# How will Language Modelers like ChatGPT Affect Occupations and Industries?

Ed Felten (Princeton)
Manav Raj (University of Pennsylvania)
Robert Seamans (New York University)[1]

18 March 2023

**Abstract**: Recent dramatic increases in AI language modeling capabilities has led to many questions about the effect of these technologies on the economy. In this paper we present a methodology to systematically assess the extent to which occupations, industries and geographies are exposed to advances in AI language modeling capabilities. We find that the top occupations exposed to language modeling include telemarketers and a variety of post-secondary teachers such as English language and literature, foreign language and literature, and history teachers. We find the top industries exposed to advances in language modeling are legal services and securities, commodities, and investments. We also find a positive correlation between wages and exposure to AI language modeling.

**Keywords**: artificial intelligence, ChatGPT, language modeling, occupation, technology

---

[1] Corresponding author: Robert Seamans (rseamans@stern.nyu.edu), NYU Stern School of Business, 44 West 4th Street, KMC 7-58, New York, NY 10012 USA. Authors listed alphabetically; all authors contributed equally. All errors are our own. First draft: 1 March 2023; this draft: 18 March 2023.

## 1. Introduction

Artificial Intelligence (AI) will likely affect the economy in many ways, potentially boosting economic growth and changing the way people work and play. The effect of AI on work will likely be multi-faceted. In some cases, AI may substitute for work previously done by humans, and in other cases AI may complement work done by humans. The effect on work will likely also vary across industries. Recent research by Goldfarb et al (2020) document that adoption of AI is relatively high in some industries such as information technology and finance, but low in others such as health care and construction. Moreover, trying to understand how AI will affect work is like trying to hit a moving target because the capabilities of AI are still advancing.

A prominent example of how AI capabilities continue to advance are the recent improvements in AI language modeling. In particular, ChatGPT, a language modeler released by Open AI in late 2022, has garnered a huge amount of attention and controversy. Some worry about the negative effects of tools like ChatGPT on jobs, as in the *New York Post* article headlined "ChatGPT could make these jobs obsolete: 'The wolf is at the door.'"[2] Others see practical and commercial promise from language modeling. For example, Microsoft announced a $10 billion partnership with Open AI and has linked ChatGPT with its Bing search engine.[3] Google felt compelled to demonstrate its own language modeler, Bard, but mistakes during the demonstration led Google's stock price to drop 7%.[4] ChatGPT has been banned by J.P. Morgan.[5] However, at present, most of this is speculation.

In order to better understand how language modelers such as ChatGPT will affect occupations, industries and geographies, we use a methodology developed by Felten et al (2018, 2021). Felten et al created the AI Occupational Exposure (AIOE) measure and used this measure to identify which occupations, industries and geographies are most exposed to AI. In this paper, we describe how the AIOE approach can be adapted to account for the recent advancement of language modeling.

---

[2] https://nypost.com/2023/01/25/chat-gpt-could-make-these-jobs-obsolete/
[3] https://www.bloomberg.com/news/articles/2023-01-23/microsoft-makes-multibillion-dollar-investment-in-openai
[4] https://www.cnbc.com/2023/02/08/alphabet-shares-slip-following-googles-ai-event-.html
[5] https://www.cbsnews.com/news/chatgpt-jpmorgan-chase-bars-workers-from-using-ai-tool/

We find that the top occupations affected include telemarketers and a variety of post-secondary teachers such as English language and literature, foreign language and literature, and history teachers. We also find the top industries exposed to advances in language modeling are legal services and securities, commodities, and investments. We also find a positive and statistically significant correlation between an occupation's mean or median wage and our measure of exposure to AI language modeling.

This article contributes to several literatures. First, by providing a systematic examination of the effect of language modeling across occupations, industries and geographies, it contributes to a nascent literature on the effects of ChatGPT and other language modelers on the economy (e.g. Agarwal et al., 2022; Zarifhonarvar, 2023). More generally, the article builds on a broader set of literature studying the effect of AI on the economy (Furman and Seamans, 2019; Goldfarb et al., 2019). Second, the article builds on and extends a set of papers that provide systematic methodologies for studying how AI affects occupations (e.g., Brynjolfsson et al, 2018; Frey & Osborne, 2017; Tolan et al., 2021; Webb, 2020). The article specifically builds off and extends the methodology described in Felten et al. (2018, 2021). In so doing, the article demonstrates the flexibility of the original Felten et al methodology; it can be adjusted dynamically to assess the impact of changes in AI capabilities. Finally, the article adds to a large literature on the effect of automating technologies on labor (e.g., Acemoglu et al., 2022; Autor, 2015; Frank et al., 2019; Genz et al., 2021).

The article proceeds as follows. Section 2 describes the AI Occupational Exposure (AIOE) measure developed by Felten et al (2018, 2021). Section 3 extends the AIOE to account for recent advances in language modeling. Section 4 provides results, including listing the top 20 most affected occupations and industries. Section 5 investigates the relationship between occupational wages and exposure to AI language modeling. Section 6 concludes.

**2. AI Occupational Exposure Methodology**

According to Felten et al (2021), the AI Occupational Exposure (AIOE) is a measure of each occupation's "exposure" to AI. The term "exposure" is used so as to be agnostic as to the effects of AI on the occupation, which could involve substitution or augmentation depending on various factors associated with the occupation itself.

The AIOE measure was constructed by linking 10 AI applications (abstract strategy games, real-time video games, image recognition, visual question answering, image generation, reading comprehension, language modeling, translation, speech recognition, and instrumental track recognition) to 52 human abilities (e.g., oral comprehension, oral expression, inductive reasoning, arm-hand steadiness, etc) using a crowd-sourced matrix that indicates the level of relatedness between each AI application and human ability. Data on the AI applications come from the Electronic Frontier Foundation (EFF) which collects and maintains statistics about the progress of AI across multiple applications. Data on human abilities comes from the Occupational Information Network (O*NET) database developed by the United States Department of Labor. O*NET uses these 52 human abilities to describe the occupational makeup of each of 800+ occupations that it tracks. Each of 800+ occupations can be thought of as a weighted combination of the 52 human abilities. O*NET uses two sets of weights: prevalence and importance.

Once the 10 AI categories and 52 human abilities are linked through the matrix, the AIOE can then be calculated for each occupation. To do this, first we calculate an ability-level exposure as follows:

$$A_{ij} = \sum_{i=1}^{10} x_{ij} \qquad (1)$$

Where *i* indexes the AI application and *j* indexes the occupational ability. The ability-level exposure, *A*, is calculated as the sum of the 10 application-ability relatedness scores, *x*, as constructed using the matrix of crowd-sourced survey data.

We then calculate the AIOE for each occupation *k* as follows:

$$AIOE_k = \frac{\sum_{j=1}^{52} A_{ij} \times L_{jk} \times I_{jk}}{\sum_{j=1}^{52} L_{jk} \times I_{jk}} \qquad (2)$$

In this equation, *i* indexes the AI application, *j* indexes the occupational ability, and *k* indexes the occupation. $A_{ij}$ represents the ability-level exposure score. We weight the ability-level AI exposure by the ability's prevalence ($L_{jk}$) and importance ($I_{jk}$) within each occupation as measured by O*NET by multiplying the ability-level AI exposure by the prevalence and importance scores for that ability within each occupation, scaled so that they are equally weighted.

Felten et al (2021) explain the construction of the AIOE scores in more detail, describe how they can be weighted at the industry level to construct an AI Industry Exposure score, or weighted at the geographic level to construct an AI Geographic Exposure score. They also provide results from a number of validation exercises and describe a number of ways in which the scores can be used by scholars and practitioners.[6]

## 3. Language Modeling AI Occupational Exposure

The original AIOE described in Felten et al (2021) explicitly weighted each of the AI applications the same. In order to update the AI Occupational Exposure score to account for advances in Language Modeling we modify equation (1) as follows.

$$A_{ij} = \sum_{i=1}^{10} \alpha_i x_{ij} \tag{3}$$

Where *i* indexes the AI application and *j* indexes the occupational ability. The ability-level exposure, *A*, is calculated as the weighted sum of the 10 application-ability relatedness scores, *x*, as constructed using the matrix of crowd-sourced survey data. $\alpha_i$ is the weight placed on each application *i*. The weights used in Felten et al (2021) set $\alpha_i$ equal to 1 for each application *i*.

Next, we set $\alpha_i$ equal to 0 for every AI application except for language modeling, which retains a weight of 1. This then constructs an ability-level exposure measure that only "counts" the value of abilities that are related to language modeling. We then proceed to calculate the $AIOE_k$ for each occupation k using this new "language modeling" weighted $A_{ij}$. The resulting $AIOE_k$ therefore captures the extent to which each occupation is exposed to advances in language modeling due to AI. A complete list of the occupations and their resulting AIOE language modeling score are listed in the appendix and on GitHub.

The resulting scores are highly correlated with the original AIOE scores (correlation coefficient: 0.979). This can be seen in Figure 1 which plots the original AIOE score and the new language modeling adjusted AIOE score for each occupation.

---

[6] The Felten et al (2021) paper is open access and available here: https://onlinelibrary.wiley.com/doi/full/10.1002/smj.3286 The data and code used to create the AIOE scores described in Felten et al (2021) is available on GitHub: https://github.com/AIOE-Data/AIOE

## 4. Results

In this section we present and briefly discuss tables of "top 20" occupations and industries exposed to language modeling.

### *4.1. Top 20 Occupations Exposed to Language Modeling*

Table 1 provides the list of top 20 occupations exposed to AI based on the original Felten et al (2021) AI Occupational Exposure (AIOE) measure as well as the top 20 occupations exposed to AI enabled advances in language modeling capabilities.

Some occupations occur in both lists, including "clinical, counseling, and school psychologists" and "history teachers, postsecondary". Notably, the language modeling list includes more education-related occupations, indicating that occupations in the field of education are likely to be relatively more impacted by advances in language modeling than other occupations. This accords well with the recent spate of articles around how ChatGPT and other language modeling tools affect the way teachers assign work and detect cheating or could use language modeling tools to develop teaching materials.

Also of interest, the top occupation in the language modeling list is "telemarketer." One might imagine that human telemarketers could benefit from language modeling being used to augment their work. For example, customer responses can be fed into a language modeling engine in real time and relevant, customer-specific prompts quickly fed to the telemarketer. Or, one might imagine that human telemarketers are substituted with language modeling enabled bots. The potential for language modeling to augment or substitute for human telemarketers work highlights one aspect of the AIOE measure: it measures "exposure" to AI, but whether that exposure leads to augmentation or substitution will depend on specifics of any given occupation.

### *4.2. Top 20 Industries Exposed to Language Modeling*

Table 2 provides the list of 20 industries most exposed to AI based on the original Felten et al. (2021) AI Industry Exposure (AIIE) measure as well as the top 20 industries exposed to AI enabled advances in language modeling capabilities.

As before, we see some similarities in the industries categorized as most exposed to AI based on the original AIOE as well as the version that focuses on advances in language modeling capabilities. For example, "Securities, Commodity Contracts, and Other Financial Investments and Related Activities" is categorized as the most exposed industry using the original AIOE and is the second most exposed industry using the language modeling-focused version of the AIOE. Legal services, insurance and employee benefit funds, and agencies, brokerages, and other insurance related activities are among the top five most exposed industries across both lists.

However, some differences emerge. One salient difference is that the language modeling-focused AIOE suggests a higher exposure to advances in AI within higher education and higher education-adjacent industries. Junior colleges, grantmaking and giving services, and business schools and computer and management training all appear within the top twenty exposed industries.

## 5. Relationship to Wages

We next investigate the relationship between two measures of wages and our language modeling AIOE. We obtain mean and median occupational wage data from the Bureau of Labor Statistics (BLS), for 2021, the most recent year for which national level occupational wage data is available.[7] Due to changes in occupational definitions over time, we do not have the same set of occupations between the 2021 BLS data and our AIOE dataset. To address this difference, we match our AIOE dataset to the BLS wage dataset first on occupational codes and then on occupational titles (which allows us to match occupations where the code has changed but the occupation is the same). We have 708 occupations present in both datasets.

To investigate the relationship between wages and exposure to AI language modeling we group occupations into 20 equal-sized bins by their language modeling AIOE score. We then plot the average AIOE score for each of these bins to the mean and median wage for the occupations in each bin. The results are plotted in Figure 2 (mean wages) and Figure 3 (median wages). The results are consistent across both figures. There appears to be a strong positive correlation between our language modeling AIOE score and mean or median wages in an occupation. In other words, occupations with higher wages are more likely to be exposed to rapid advances in language modeling from products such as ChatGPT or others. Given that the most exposed occupations appear to be white-collar occupations that may be likely to be classified as "high-skilled" labor, this is perhaps not surprising.

---

[7] https://www.bls.gov/oes/

## 6. Conclusion

In this paper we present a methodology to systematically assess the extent to which occupations and industries are exposed to advances in AI language modeling capabilities. This methodology relies on the approach described in Felten et al (2021) but adapts it to account for recent advances in language modeling. We find that the top occupations exposed to language modeling include telemarketers and a variety of post-secondary teachers such as English language and literature, foreign language and literature, and history teachers. We also find the top industries exposed to advances in language modeling are legal services and securities, commodities, and investments. We also find that occupations with higher wages are more likely to be exposed to rapid advances in language modeling.

At a broad level, this paper adds to a growing literature studying the effects of AI on labor and work. More specifically, the paper provides a systematic approach for understanding how ChatGPT and other language modelers will affect occupations, industries and geographies. We believe these results will be useful for other scholars as well as practitioners and policymakers.

**Figure 1: Comparison between Original AIOE and Language Modeling Adjusted AIOE**

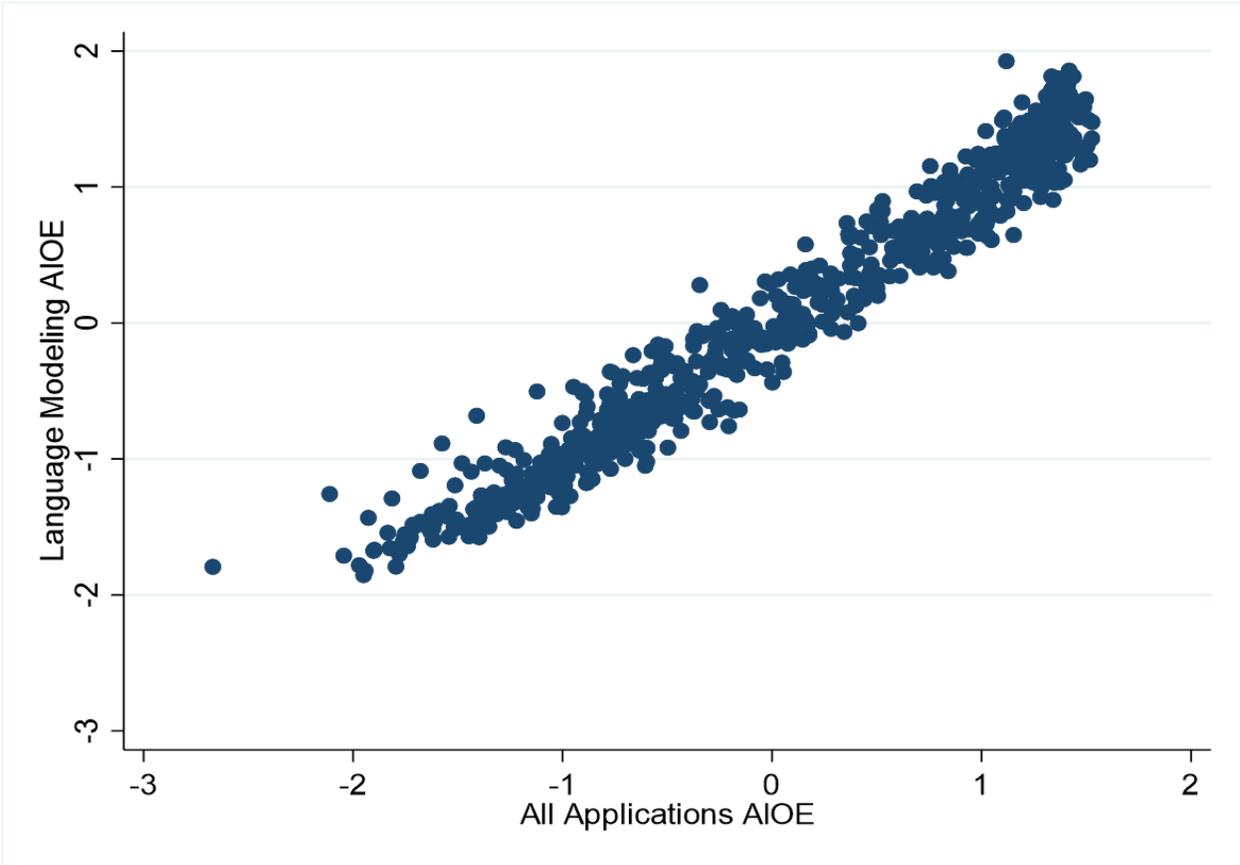

Notes: This figure plots the original AIOE score (x-axis) and the new language modeling adjusted AIOE score (y-axis) for each occupation.

**Figure 2: Relationship between Language Modeling AIOE and Mean Occupational Wages**

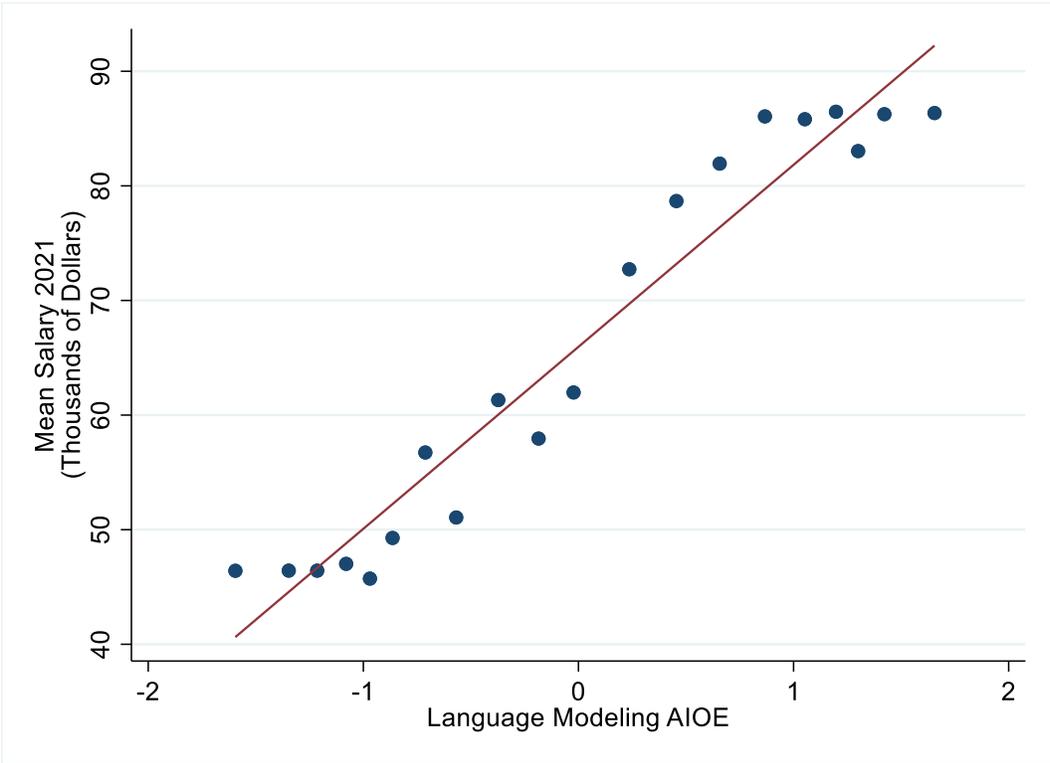

Notes: This figure plots the relationship between the language modeling AIOE score (x-axis) and mean salary (y-axis) for each occupation, where occupations are grouped into 20 equal sized bins based on language modeling AIOE score.

**Figure 3: Relationship between Language Modeling AIOE and Median Occupational Wages**

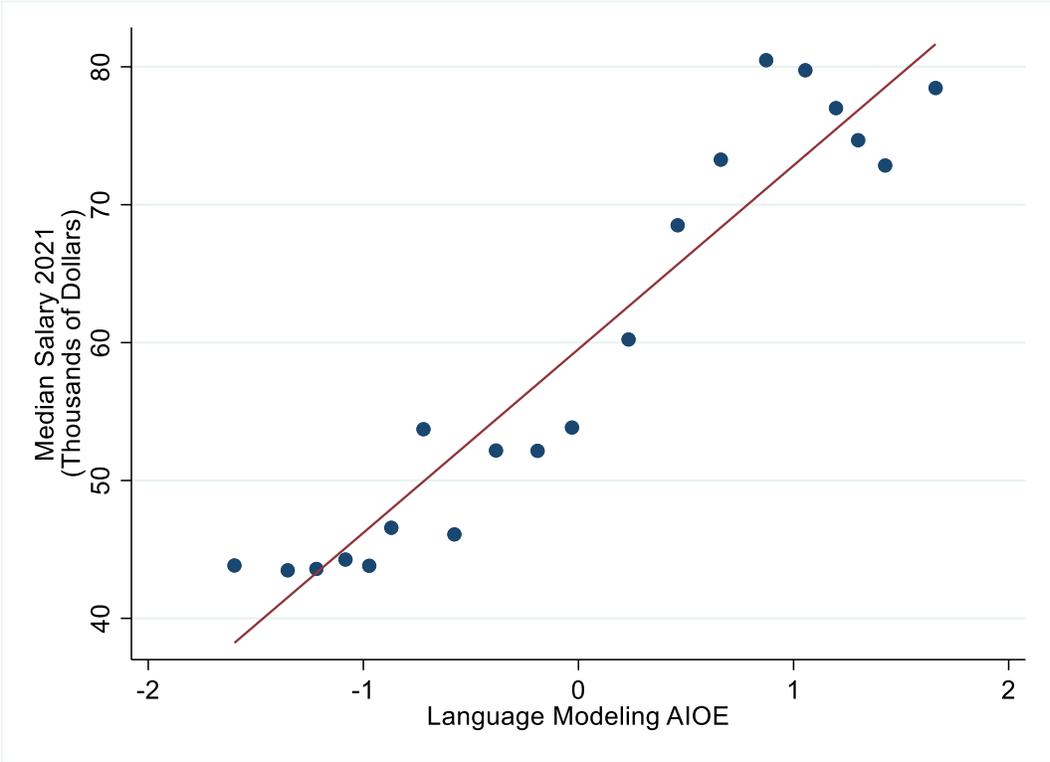

Notes: This figure plots the relationship between the language modeling AIOE score (x-axis) and median salary (y-axis) for each occupation, where occupations are grouped into 20 equal sized bins based on language modeling AIOE score.

**Table 1: Top 20 Occupations Exposed to AI, Original and with Language Modeling Adjustment**

| Rank | Top 20 Occupations from Original AIOE | Top 20 Occupations after Language Modeling Adjustment |
|---|---|---|
| 1 | Genetic Counselors | Telemarketers |
| 2 | Financial Examiners | English Language and Literature Teachers, Postsecondary |
| 3 | Actuaries | Foreign Language and Literature Teachers, Postsecondary |
| 4 | Purchasing Agents, Except Wholesale, Retail, and Farm Products | History Teachers, Postsecondary |
| 5 | Budget Analysts | Law Teachers, Postsecondary |
| 6 | Judges, Magistrate Judges, and Magistrates | Philosophy and Religion Teachers, Postsecondary |
| 7 | Procurement Clerks | Sociology Teachers, Postsecondary |
| 8 | Accountants and Auditors | Political Science Teachers, Postsecondary |
| 9 | Mathematicians | Criminal Justice and Law Enforcement Teachers, Postsecondary |
| 10 | Judicial Law Clerks | Sociologists |
| 11 | Education Administrators, Postsecondary | Social Work Teachers, Postsecondary |
| 12 | Clinical, Counseling, and School Psychologists | Psychology Teachers, Postsecondary |
| 13 | Financial Managers | Communications Teachers, Postsecondary |
| 14 | Compensation, Benefits, and Job Analysis Specialists | Political Scientists |
| 15 | Credit Authorizers, Checkers, and Clerks | Area, Ethnic, and Cultural Studies Teachers, Postsecondary |
| 16 | History Teachers, Postsecondary | Arbitrators, Mediators, and Conciliators |
| 17 | Geographers | Judges, Magistrate Judges, and Magistrates |
| 18 | Epidemiologists | Geography Teachers, Postsecondary |
| 19 | Management Analysts | Library Science Teachers, Postsecondary |
| 20 | Arbitrators, Mediators, and Conciliators | Clinical, Counseling, and School Psychologists |

Notes: This table lists the top 20 occupations most exposed to AI from the original AIOE (Felten et al., 2021) and the top 20 occupations most exposed to advances in AI language modeling.

**Table 2: Top 20 Industries Exposed to AI, Original and with Language Modeling Adjustment**

| Rank | Top 20 Industries from Original AIOE | Top 20 Industries after Language Modeling Adjustment |
|---|---|---|
| 1 | Securities, Commodity Contracts, and Other Financial Investments and Related Activities | Legal Services |
| 2 | Accounting, Tax Preparation, Bookkeeping, and Payroll Services | Securities, Commodity Contracts, and Other Financial Investments and Related Activities |
| 3 | Insurance and Employee Benefit Funds | Agencies, Brokerages, and Other Insurance Related Activities |
| 4 | Legal Services | Insurance and Employee Benefit Funds |
| 5 | Agencies, Brokerages, and Other Insurance Related Activities | Nondepository Credit Intermediation |
| 6 | Nondepository Credit Intermediation | Agents and Managers for Artists, Athletes, Entertainers, and Other Public Figures |
| 7 | Other Investment Pools and Funds | Insurance Carriers |
| 8 | Insurance Carriers | Other Investment Pools and Funds |
| 9 | Software Publishers | Accounting, Tax Preparation, Bookkeeping, and Payroll Services |
| 10 | Lessors of Nonfinancial Intangible Assets (except Copyrighted Works) | Business Support Services |
| 11 | Agents and Managers for Artists, Athletes, Entertainers, and Other Public Figures | Software Publishers |
| 12 | Credit Intermediation and Related Activities (5221 And 5223 only) | Lessors of Nonfinancial Intangible Assets (except Copyrighted Works) |
| 13 | Computer Systems Design and Related Services | Business Schools and Computer and Management Training |
| 14 | Management, Scientific, and Technical Consulting Services | Credit Intermediation and Related Activities (5221 And 5223 only) |
| 15 | Monetary Authorities-Central Bank | Grantmaking and Giving Services |
| 16 | Office Administrative Services | Travel Arrangement and Reservation Services |
| 17 | Other Information Services | Junior Colleges |
| 18 | Data Processing, Hosting, and Related Services | Computer Systems Design and Related Services |
| 19 | Business Schools and Computer and Management Training | Management, Scientific, and Technical Consulting Services |
| 20 | Grantmaking and Giving Services | Other Information Services |

Notes: This table lists the top 20 industries most exposed to AI from the original AIOE (Felten et al., 2021) and the top 20 industries most exposed to advances in AI language modeling.

# APPENDIX

**FULL LIST OF OCCUPATIONS
SORTED BY LANGUAGE MODELING EXPOSURE SCORE**

| Rank | SOC Code | Occupation Title | Language Modeling AIOE |
|---|---|---|---|
| 1 | 41-9041 | Telemarketers | 1.926 |
| 2 | 25-1123 | English Language and Literature Teachers, Postsecondary | 1.857 |
| 3 | 25-1124 | Foreign Language and Literature Teachers, Postsecondary | 1.814 |
| 4 | 25-1125 | History Teachers, Postsecondary | 1.813 |
| 5 | 25-1112 | Law Teachers, Postsecondary | 1.802 |
| 6 | 25-1126 | Philosophy and Religion Teachers, Postsecondary | 1.800 |
| 7 | 25-1067 | Sociology Teachers, Postsecondary | 1.770 |
| 8 | 25-1065 | Political Science Teachers, Postsecondary | 1.770 |
| 9 | 25-1111 | Criminal Justice and Law Enforcement Teachers, Postsecondary | 1.754 |
| 10 | 19-3041 | Sociologists | 1.747 |
| 11 | 25-1113 | Social Work Teachers, Postsecondary | 1.739 |
| 12 | 25-1066 | Psychology Teachers, Postsecondary | 1.716 |
| 13 | 25-1122 | Communications Teachers, Postsecondary | 1.702 |
| 14 | 19-3094 | Political Scientists | 1.687 |
| 15 | 25-1062 | Area, Ethnic, and Cultural Studies Teachers, Postsecondary | 1.669 |
| 16 | 23-1022 | Arbitrators, Mediators, and Conciliators | 1.647 |
| 17 | 23-1023 | Judges, Magistrate Judges, and Magistrates | 1.646 |
| 18 | 25-1064 | Geography Teachers, Postsecondary | 1.629 |
| 19 | 25-1082 | Library Science Teachers, Postsecondary | 1.626 |
| 20 | 19-3031 | Clinical, Counseling, and School Psychologists | 1.626 |
| 21 | 25-1081 | Education Teachers, Postsecondary | 1.624 |
| 22 | 25-1011 | Business Teachers, Postsecondary | 1.618 |
| 23 | 25-1053 | Environmental Science Teachers, Postsecondary | 1.603 |
| 24 | 43-3061 | Procurement Clerks | 1.590 |
| 25 | 25-1043 | Forestry and Conservation Science Teachers, Postsecondary | 1.563 |
| 26 | 13-1071 | Human Resources Specialists | 1.557 |
| 27 | 13-1111 | Management Analysts | 1.548 |
| 28 | 23-1021 | Administrative Law Judges, Adjudicators, and Hearing Officers | 1.547 |
| 29 | 43-4041 | Credit Authorizers, Checkers, and Clerks | 1.546 |
| 30 | 11-9033 | Education Administrators, Postsecondary | 1.545 |
| 31 | 21-1014 | Mental Health Counselors | 1.537 |
| 32 | 25-1061 | Anthropology and Archeology Teachers, Postsecondary | 1.534 |
| 33 | 25-1051 | Atmospheric, Earth, Marine, and Space Sciences Teachers, Postsecondary | 1.522 |
| 34 | 27-3031 | Public Relations Specialists | 1.518 |
| 35 | 23-1012 | Judicial Law Clerks | 1.513 |
| 36 | 41-9091 | Door-To-Door Sales Workers, News and Street Vendors, and Related Workers | 1.511 |
| 37 | 13-1023 | Purchasing Agents, Except Wholesale, Retail, and Farm Products | 1.496 |
| 38 | 25-1042 | Biological Science Teachers, Postsecondary | 1.493 |

| 39 | 19-3093 | Historians | 1.491 |
|---|---|---|---|
| 40 | 25-3099 | Teachers and Instructors, All Other | 1.486 |
| 41 | 43-4161 | Human Resources Assistants, Except Payroll and Timekeeping | 1.481 |
| 42 | 29-9092 | Genetic Counselors | 1.478 |
| 43 | 13-2072 | Loan Officers | 1.477 |
| 44 | 21-1013 | Marriage and Family Therapists | 1.473 |
| 45 | 27-3022 | Reporters and Correspondents | 1.471 |
| 46 | 21-2011 | Clergy | 1.470 |
| 47 | 25-1192 | Home Economics Teachers, Postsecondary | 1.468 |
| 48 | 11-3111 | Compensation and Benefits Managers | 1.467 |
| 49 | 25-1063 | Economics Teachers, Postsecondary | 1.462 |
| 50 | 23-1011 | Lawyers | 1.454 |
| 51 | 25-1052 | Chemistry Teachers, Postsecondary | 1.450 |
| 52 | 43-4111 | Interviewers, Except Eligibility and Loan | 1.449 |
| 53 | 25-3011 | Adult Basic and Secondary Education and Literacy Teachers and Instructors | 1.440 |
| 54 | 43-9081 | Proofreaders and Copy Markers | 1.436 |
| 55 | 13-1075 | Labor Relations Specialists | 1.431 |
| 56 | 25-1193 | Recreation and Fitness Studies Teachers, Postsecondary | 1.429 |
| 57 | 19-3032 | Industrial-Organizational Psychologists | 1.427 |
| 58 | 41-3021 | Insurance Sales Agents | 1.427 |
| 59 | 43-4061 | Eligibility Interviewers, Government Programs | 1.417 |
| 60 | 43-2021 | Telephone Operators | 1.412 |
| 61 | 25-1022 | Mathematical Science Teachers, Postsecondary | 1.407 |
| 62 | 13-1131 | Fundraisers | 1.405 |
| 63 | 13-2052 | Personal Financial Advisors | 1.396 |
| 64 | 11-3061 | Purchasing Managers | 1.393 |
| 65 | 25-1021 | Computer Science Teachers, Postsecondary | 1.390 |
| 66 | 23-2093 | Title Examiners, Abstractors, and Searchers | 1.384 |
| 67 | 11-2031 | Public Relations and Fundraising Managers | 1.383 |
| 68 | 25-1191 | Graduate Teaching Assistants | 1.376 |
| 69 | 29-1066 | Psychiatrists | 1.366 |
| 70 | 29-1031 | Dietitians and Nutritionists | 1.364 |
| 71 | 19-3092 | Geographers | 1.363 |
| 72 | 13-1141 | Compensation, Benefits, and Job Analysis Specialists | 1.362 |
| 73 | 11-3121 | Human Resources Managers | 1.359 |
| 74 | 13-2061 | Financial Examiners | 1.358 |
| 75 | 13-1161 | Market Research Analysts and Marketing Specialists | 1.356 |
| 76 | 13-2071 | Credit Counselors | 1.355 |
| 77 | 21-2021 | Directors, Religious Activities and Education | 1.355 |
| 78 | 27-3091 | Interpreters and Translators | 1.352 |
| 79 | 41-9021 | Real Estate Brokers | 1.348 |
| 80 | 29-1127 | Speech-Language Pathologists | 1.343 |

| 81 | 25-1032 | Engineering Teachers, Postsecondary | 1.337 |
|---|---|---|---|
| 82 | 11-3131 | Training and Development Managers | 1.337 |
| 83 | 43-4031 | Court, Municipal, and License Clerks | 1.336 |
| 84 | 25-2053 | Special Education Teachers, Middle School | 1.334 |
| 85 | 23-2011 | Paralegals and Legal Assistants | 1.333 |
| 86 | 19-3011 | Economists | 1.330 |
| 87 | 19-3039 | Psychologists, All Other | 1.330 |
| 88 | 25-9031 | Instructional Coordinators | 1.324 |
| 89 | 13-1151 | Training and Development Specialists | 1.323 |
| 90 | 27-3042 | Technical Writers | 1.317 |
| 91 | 11-9111 | Medical and Health Services Managers | 1.315 |
| 92 | 11-1011 | Chief Executives | 1.309 |
| 93 | 13-2031 | Budget Analysts | 1.300 |
| 94 | 43-6013 | Medical Secretaries | 1.298 |
| 95 | 11-3031 | Financial Managers | 1.295 |
| 96 | 11-2022 | Sales Managers | 1.294 |
| 97 | 25-2022 | Middle School Teachers, Except Special and Career/Technical Education | 1.284 |
| 98 | 19-1041 | Epidemiologists | 1.284 |
| 99 | 19-3022 | Survey Researchers | 1.279 |
| 100 | 41-3041 | Travel Agents | 1.278 |
| 101 | 43-4131 | Loan Interviewers and Clerks | 1.274 |
| 102 | 13-2051 | Financial Analysts | 1.273 |
| 103 | 43-3011 | Bill and Account Collectors | 1.270 |
| 104 | 13-2053 | Insurance Underwriters | 1.262 |
| 105 | 41-3031 | Securities, Commodities, and Financial Services Sales Agents | 1.260 |
| 106 | 19-2021 | Atmospheric and Space Scientists | 1.259 |
| 107 | 41-1012 | First-Line Supervisors of Non-Retail Sales Workers | 1.259 |
| 108 | 43-9041 | Insurance Claims and Policy Processing Clerks | 1.257 |
| 109 | 43-5011 | Cargo and Freight Agents | 1.255 |
| 110 | 43-6014 | Secretaries and Administrative Assistants, Except Legal, Medical, and Executive | 1.254 |
| 111 | 13-2081 | Tax Examiners and Collectors, and Revenue Agents | 1.253 |
| 112 | 21-1023 | Mental Health and Substance Abuse Social Workers | 1.253 |
| 113 | 13-2099 | Financial Specialists, All Other | 1.253 |
| 114 | 25-2031 | Secondary School Teachers, Except Special and Career/Technical Education | 1.249 |
| 115 | 13-1011 | Agents and Business Managers of Artists, Performers, and Athletes | 1.248 |
| 116 | 25-1031 | Architecture Teachers, Postsecondary | 1.246 |
| 117 | 43-2011 | Switchboard Operators, Including Answering Service | 1.244 |
| 118 | 15-2041 | Statisticians | 1.243 |
| 119 | 41-3099 | Sales Representatives, Services, All Other | 1.242 |

| 120 | 11-9121 | Natural Sciences Managers | 1.241 |
|---|---|---|---|
| 121 | 25-1054 | Physics Teachers, Postsecondary | 1.240 |
| 122 | 15-2031 | Operations Research Analysts | 1.234 |
| 123 | 27-3021 | Broadcast News Analysts | 1.230 |
| 124 | 25-1071 | Health Specialties Teachers, Postsecondary | 1.226 |
| 125 | 11-9151 | Social and Community Service Managers | 1.225 |
| 126 | 11-2011 | Advertising and Promotions Managers | 1.217 |
| 127 | 11-9141 | Property, Real Estate, and Community Association Managers | 1.215 |
| 128 | 13-2011 | Accountants and Auditors | 1.214 |
| 129 | 21-1011 | Substance Abuse and Behavioral Disorder Counselors | 1.204 |
| 130 | 11-2021 | Marketing Managers | 1.204 |
| 131 | 43-4141 | New Accounts Clerks | 1.201 |
| 132 | 15-2011 | Actuaries | 1.200 |
| 133 | 11-9032 | Education Administrators, Elementary and Secondary School | 1.199 |
| 134 | 21-1012 | Educational, Guidance, School, and Vocational Counselors | 1.196 |
| 135 | 31-9094 | Medical Transcriptionists | 1.196 |
| 136 | 27-3041 | Editors | 1.191 |
| 137 | 43-4021 | Correspondence Clerks | 1.184 |
| 138 | 27-3043 | Writers and Authors | 1.170 |
| 139 | 19-1042 | Medical Scientists, Except Epidemiologists | 1.167 |
| 140 | 15-2021 | Mathematicians | 1.166 |
| 141 | 15-1133 | Software Developers, Systems Software | 1.166 |
| 142 | 43-3051 | Payroll and Timekeeping Clerks | 1.155 |
| 143 | 25-1041 | Agricultural Sciences Teachers, Postsecondary | 1.154 |
| 144 | 43-6012 | Legal Secretaries | 1.149 |
| 145 | 29-1065 | Pediatricians, General | 1.146 |
| 146 | 13-2082 | Tax Preparers | 1.144 |
| 147 | 43-3021 | Billing and Posting Clerks | 1.136 |
| 148 | 43-1011 | First-Line Supervisors of Office and Administrative Support Workers | 1.134 |
| 149 | 13-1051 | Cost Estimators | 1.134 |
| 150 | 43-5032 | Dispatchers, Except Police, Fire, and Ambulance | 1.128 |
| 151 | 19-4061 | Social Science Research Assistants | 1.126 |
| 152 | 27-3011 | Radio and Television Announcers | 1.124 |
| 153 | 41-3011 | Advertising Sales Agents | 1.122 |
| 154 | 17-2011 | Aerospace Engineers | 1.117 |
| 155 | 29-1062 | Family and General Practitioners | 1.115 |
| 156 | 13-1031 | Claims Adjusters, Examiners, and Investigators | 1.106 |
| 157 | 43-6011 | Executive Secretaries and Executive Administrative Assistants | 1.106 |
| 158 | 19-2012 | Physicists | 1.097 |
| 159 | 21-1022 | Healthcare Social Workers | 1.094 |
| 160 | 13-2041 | Credit Analysts | 1.092 |
| 161 | 19-3099 | Social Scientists and Related Workers, All Other | 1.085 |
| 162 | 13-1081 | Logisticians | 1.081 |
| 163 | 43-4051 | Customer Service Representatives | 1.076 |
| 164 | 19-3051 | Urban and Regional Planners | 1.075 |
| 165 | 17-2161 | Nuclear Engineers | 1.073 |
| 166 | 39-6012 | Concierges | 1.071 |
| 167 | 43-9111 | Statistical Assistants | 1.053 |
| 168 | 15-1121 | Computer Systems Analysts | 1.046 |
| 169 | 15-1122 | Information Security Analysts | 1.045 |
| 170 | 17-2081 | Environmental Engineers | 1.044 |
| 171 | 15-1199 | Computer Occupations, All Other | 1.044 |
| 172 | 41-4012 | Sales Representatives, Wholesale and Manufacturing, Except Technical and Scientific Products | 1.039 |
| 173 | 19-2011 | Astronomers | 1.035 |
| 174 | 17-2171 | Petroleum Engineers | 1.033 |
| 175 | 43-4171 | Receptionists and Information Clerks | 1.033 |
| 176 | 27-2041 | Music Directors and Composers | 1.032 |
| 177 | 17-2151 | Mining and Geological Engineers, Including Mining Safety Engineers | 1.031 |
| 178 | 43-4011 | Brokerage Clerks | 1.026 |
| 179 | 41-9031 | Sales Engineers | 1.026 |
| 180 | 15-1131 | Computer Programmers | 1.025 |
| 181 | 25-2021 | Elementary School Teachers, Except Special Education | 1.017 |
| 182 | 39-1011 | Gaming Supervisors | 1.015 |
| 183 | 25-1121 | Art, Drama, and Music Teachers, Postsecondary | 1.007 |
| 184 | 29-1063 | Internists, General | 1.001 |
| 185 | 25-2054 | Special Education Teachers, Secondary School | 1.000 |
| 186 | 15-1141 | Database Administrators | 0.996 |
| 187 | 25-4021 | Librarians | 0.978 |
| 188 | 17-2061 | Computer Hardware Engineers | 0.977 |
| 189 | 19-1011 | Animal Scientists | 0.972 |
| 190 | 21-1091 | Health Educators | 0.968 |
| 191 | 15-1111 | Computer and Information Research Scientists | 0.968 |
| 192 | 11-9031 | Education Administrators, Preschool and Childcare Center/Program | 0.967 |
| 193 | 21-1021 | Child, Family, and School Social Workers | 0.957 |
| 194 | 41-4011 | Sales Representatives, Wholesale and Manufacturing, Technical and Scientific Products | 0.953 |
| 195 | 11-3021 | Computer and Information Systems Managers | 0.945 |
| 196 | 27-4013 | Radio Operators | 0.941 |
| 197 | 25-2052 | Special Education Teachers, Kindergarten and Elementary School | 0.937 |
| 198 | 17-2051 | Civil Engineers | 0.927 |
| 199 | 13-1199 | Business Operations Specialists, All Other | 0.925 |
| 200 | 19-2099 | Physical Scientists, All Other | 0.913 |
| 201 | 43-9061 | Office Clerks, General | 0.911 |
| 202 | 11-9199 | Managers, All Other | 0.910 |
| 203 | 17-2041 | Chemical Engineers | 0.906 |
| 204 | 17-2112 | Industrial Engineers | 0.905 |
| 205 | 11-3071 | Transportation, Storage, and Distribution Managers | 0.905 |
| 206 | 23-2091 | Court Reporters | 0.903 |
| 207 | 17-2071 | Electrical Engineers | 0.901 |
| 208 | 11-9161 | Emergency Management Directors | 0.901 |
| 209 | 21-1092 | Probation Officers and Correctional Treatment Specialists | 0.897 |
| 210 | 19-1029 | Biological Scientists, All Other | 0.883 |
| 211 | 15-1132 | Software Developers, Applications | 0.882 |
| 212 | 19-2041 | Environmental Scientists and Specialists, Including Health | 0.870 |
| 213 | 11-9131 | Postmasters and Mail Superintendents | 0.866 |
| 214 | 27-2012 | Producers and Directors | 0.860 |
| 215 | 17-1012 | Landscape Architects | 0.836 |
| 216 | 21-1015 | Rehabilitation Counselors | 0.835 |
| 217 | 25-3021 | Self-Enrichment Education Teachers | 0.825 |
| 218 | 27-1025 | Interior Designers | 0.822 |
| 219 | 43-3031 | Bookkeeping, Accounting, and Auditing Clerks | 0.818 |
| 220 | 19-1021 | Biochemists and Biophysicists | 0.794 |
| 221 | 13-1022 | Wholesale and Retail Buyers, Except Farm Products | 0.791 |
| 222 | 15-1134 | Web Developers | 0.789 |
| 223 | 19-2032 | Materials Scientists | 0.786 |
| 224 | 25-9021 | Farm and Home Management Advisors | 0.784 |
| 225 | 17-2131 | Materials Engineers | 0.780 |
| 226 | 43-5031 | Police, Fire, and Ambulance Dispatchers | 0.774 |
| 227 | 27-3012 | Public Address System and Other Announcers | 0.773 |
| 228 | 11-3011 | Administrative Services Managers | 0.768 |
| 229 | 17-2111 | Health and Safety Engineers, Except Mining Safety Engineers and Inspectors | 0.764 |
| 230 | 43-4151 | Order Clerks | 0.761 |
| 231 | 25-9041 | Teacher Assistants | 0.748 |
| 232 | 25-1072 | Nursing Instructors and Teachers, Postsecondary | 0.744 |
| 233 | 19-3091 | Anthropologists and Archeologists | 0.739 |
| 234 | 13-1021 | Buyers and Purchasing Agents, Farm Products | 0.737 |
| 235 | 39-7011 | Tour Guides and Escorts | 0.736 |
| 236 | 17-2031 | Biomedical Engineers | 0.734 |
| 237 | 27-1011 | Art Directors | 0.727 |
| 238 | 11-9071 | Gaming Managers | 0.719 |
| 239 | 43-4081 | Hotel, Motel, and Resort Desk Clerks | 0.711 |
| 240 | 39-9041 | Residential Advisors | 0.711 |
| 241 | 41-9022 | Real Estate Sales Agents | 0.701 |
| 242 | 29-1181 | Audiologists | 0.696 |
| 243 | 25-4011 | Archivists | 0.693 |
| 244 | 17-2021 | Agricultural Engineers | 0.690 |
| 245 | 29-1199 | Health Diagnosing and Treating Practitioners, All Other | 0.679 |
| 246 | 17-2141 | Mechanical Engineers | 0.678 |
| 247 | 11-1021 | General and Operations Managers | 0.678 |
| 248 | 17-3012 | Electrical and Electronics Drafters | 0.676 |
| 249 | 19-1020 | Biologists | 0.662 |
| 250 | 43-5061 | Production, Planning, and Expediting Clerks | 0.661 |
| 251 | 27-4032 | Film and Video Editors | 0.657 |
| 252 | 15-1152 | Computer Network Support Specialists | 0.657 |
| 253 | 25-4012 | Curators | 0.656 |
| 254 | 25-2051 | Special Education Teachers, Preschool | 0.655 |
| 255 | 25-2012 | Kindergarten Teachers, Except Special Education | 0.653 |
| 256 | 53-2021 | Air Traffic Controllers | 0.648 |
| 257 | 11-9041 | Architectural and Engineering Managers | 0.647 |
| 258 | 25-2023 | Career/Technical Education Teachers, Middle School | 0.647 |
| 259 | 11-9021 | Construction Managers | 0.636 |
| 260 | 13-2021 | Appraisers and Assessors of Real Estate | 0.628 |
| 261 | 39-1021 | First-Line Supervisors of Personal Service Workers | 0.628 |
| 262 | 11-9039 | Education Administrators, All Other | 0.628 |
| 263 | 29-1069 | Physicians and Surgeons, All Other | 0.619 |
| 264 | 33-9021 | Private Detectives and Investigators | 0.615 |
| 265 | 17-1011 | Architects, Except Landscape and Naval | 0.610 |
| 266 | 19-1013 | Soil and Plant Scientists | 0.609 |
| 267 | 29-9011 | Occupational Health and Safety Specialists | 0.598 |
| 268 | 13-1032 | Insurance Appraisers, Auto Damage | 0.582 |
| 269 | 17-2199 | Engineers, All Other | 0.581 |
| 270 | 27-2011 | Actors | 0.578 |
| 271 | 19-1012 | Food Scientists and Technologists | 0.573 |
| 272 | 13-1041 | Compliance Officers | 0.572 |
| 273 | 29-1051 | Pharmacists | 0.571 |
| 274 | 19-2042 | Geoscientists, Except Hydrologists and Geographers | 0.567 |
| 275 | 27-1021 | Commercial and Industrial Designers | 0.560 |
| 276 | 21-1093 | Social and Human Service Assistants | 0.557 |
| 277 | 19-2043 | Hydrologists | 0.555 |
| 278 | 17-3011 | Architectural and Civil Drafters | 0.554 |
| 279 | 29-1081 | Podiatrists | 0.554 |
| 280 | 17-3022 | Civil Engineering Technicians | 0.553 |
| 281 | 39-7012 | Travel Guides | 0.553 |
| 282 | 29-2071 | Medical Records and Health Information Technicians | 0.552 |
| 283 | 27-1024 | Graphic Designers | 0.527 |
| 284 | 13-1121 | Meeting, Convention, and Event Planners | 0.513 |
| 285 | 27-1022 | Fashion Designers | 0.508 |
| 286 | 43-9022 | Word Processors and Typists | 0.507 |
| 287 | 43-3041 | Gaming Cage Workers | 0.496 |
| 288 | 15-1143 | Computer Network Architects | 0.495 |
| 289 | 15-1142 | Network and Computer Systems Administrators | 0.475 |
| 290 | 19-1022 | Microbiologists | 0.474 |
| 291 | 17-1021 | Cartographers and Photogrammetrists | 0.472 |
| 292 | 11-9081 | Lodging Managers | 0.464 |
| 293 | 29-2092 | Hearing Aid Specialists | 0.458 |
| 294 | 17-2072 | Electronics Engineers, Except Computer | 0.457 |
| 295 | 15-2091 | Mathematical Technicians | 0.438 |
| 296 | 33-9031 | Gaming Surveillance Officers and Gaming Investigators | 0.433 |
| 297 | 27-1014 | Multimedia Artists and Animators | 0.432 |
| 298 | 43-3071 | Tellers | 0.431 |
| 299 | 43-4181 | Reservation and Transportation Ticket Agents and Travel Clerks | 0.423 |
| 300 | 25-2032 | Career/Technical Education Teachers, Secondary School | 0.422 |
| 301 | 51-8011 | Nuclear Power Reactor Operators | 0.409 |
| 302 | 43-9011 | Computer Operators | 0.408 |
| 303 | 29-1125 | Recreational Therapists | 0.401 |
| 304 | 25-1194 | Vocational Education Teachers, Postsecondary | 0.390 |
| 305 | 17-3013 | Mechanical Drafters | 0.382 |
| 306 | 39-3012 | Gaming and Sports Book Writers and Runners | 0.365 |
| 307 | 29-1064 | Obstetricians and Gynecologists | 0.364 |
| 308 | 41-2031 | Retail Salespersons | 0.357 |
| 309 | 11-3051 | Industrial Production Managers | 0.354 |
| 310 | 17-3026 | Industrial Engineering Technicians | 0.347 |
| 311 | 17-2121 | Marine Engineers and Naval Architects | 0.343 |
| 312 | 27-4014 | Sound Engineering Technicians | 0.338 |
| 313 | 11-9061 | Funeral Service Managers | 0.337 |
| 314 | 41-9011 | Demonstrators and Product Promoters | 0.334 |
| 315 | 29-9012 | Occupational Health and Safety Technicians | 0.329 |
| 316 | 19-1023 | Zoologists and Wildlife Biologists | 0.327 |
| 317 | 19-2031 | Chemists | 0.327 |
| 318 | 15-1151 | Computer User Support Specialists | 0.326 |
| 319 | 25-2011 | Preschool Teachers, Except Special Education | 0.322 |
| 320 | 27-2022 | Coaches and Scouts | 0.319 |
| 321 | 29-1128 | Exercise Physiologists | 0.307 |
| 322 | 29-1161 | Nurse Midwives | 0.305 |
| 323 | 29-1171 | Nurse Practitioners | 0.302 |
| 324 | 27-2042 | Musicians and Singers | 0.290 |
| 325 | 39-3093 | Locker Room, Coatroom, and Dressing Room Attendants | 0.279 |
| 326 | 29-9099 | Healthcare Practitioners and Technical Workers, All Other | 0.279 |
| 327 | 43-4121 | Library Assistants, Clerical | 0.276 |
| 328 | 29-1141 | Registered Nurses | 0.272 |

| 329 | 25-9011 | Audio-Visual and Multimedia Collections Specialists | 0.272 |
|---|---|---|---|
| 330 | 29-1122 | Occupational Therapists | 0.266 |
| 331 | 51-8012 | Power Distributors and Dispatchers | 0.257 |
| 332 | 31-9099 | Healthcare Support Workers, All Other | 0.252 |
| 333 | 53-2022 | Airfield Operations Specialists | 0.246 |
| 334 | 43-4071 | File Clerks | 0.240 |
| 335 | 31-9092 | Medical Assistants | 0.236 |
| 336 | 29-1061 | Anesthesiologists | 0.212 |
| 337 | 41-1011 | First-Line Supervisors of Retail Sales Workers | 0.201 |
| 338 | 17-3021 | Aerospace Engineering and Operations Technicians | 0.201 |
| 339 | 29-1041 | Optometrists | 0.200 |
| 340 | 35-9031 | Hosts and Hostesses, Restaurant, Lounge, and Coffee Shop | 0.182 |
| 341 | 25-4031 | Library Technicians | 0.180 |
| 342 | 19-4099 | Life, Physical, and Social Science Technicians, All Other | 0.176 |
| 343 | 43-9021 | Data Entry Keyers | 0.172 |
| 344 | 33-3021 | Detectives and Criminal Investigators | 0.151 |
| 345 | 41-2021 | Counter and Rental Clerks | 0.143 |
| 346 | 53-1021 | First-Line Supervisors of Helpers, Laborers, and Material Movers, Hand | 0.138 |
| 347 | 21-1094 | Community Health Workers | 0.135 |
| 348 | 29-1151 | Nurse Anesthetists | 0.134 |
| 349 | 19-4021 | Biological Technicians | 0.133 |
| 350 | 53-6041 | Traffic Technicians | 0.129 |
| 351 | 29-2091 | Orthotists and Prosthetists | 0.126 |
| 352 | 39-9032 | Recreation Workers | 0.097 |
| 353 | 29-2057 | Ophthalmic Medical Technicians | 0.083 |
| 354 | 17-3023 | Electrical and Electronic Engineering Technicians | 0.078 |
| 355 | 19-1031 | Conservation Scientists | 0.074 |
| 356 | 53-1031 | First-Line Supervisors of Transportation and Material-Moving Machine and Vehicle Operators | 0.068 |
| 357 | 47-4011 | Construction and Building Inspectors | 0.068 |
| 358 | 11-9051 | Food Service Managers | 0.061 |
| 359 | 29-1011 | Chiropractors | 0.052 |
| 360 | 29-1071 | Physician Assistants | 0.040 |
| 361 | 39-3011 | Gaming Dealers | 0.030 |
| 362 | 29-2081 | Opticians, Dispensing | 0.029 |
| 363 | 39-1012 | Slot Supervisors | 0.027 |
| 364 | 29-1023 | Orthodontists | 0.021 |
| 365 | 19-4092 | Forensic Science Technicians | 0.010 |
| 366 | 29-2053 | Psychiatric Technicians | -0.001 |
| 367 | 51-5111 | Prepress Technicians and Workers | -0.001 |
| 368 | 45-2011 | Agricultural Inspectors | -0.008 |
| 369 | 41-2022 | Parts Salespersons | -0.014 |
| 370 | 33-9099 | Protective Service Workers, All Other | -0.017 |
| 371 | 29-2033 | Nuclear Medicine Technologists | -0.023 |
| 372 | 29-2052 | Pharmacy Technicians | -0.023 |
| 373 | 19-4041 | Geological and Petroleum Technicians | -0.027 |
| 374 | 31-9095 | Pharmacy Aides | -0.028 |
| 375 | 51-1011 | First-Line Supervisors of Production and Operating Workers | -0.037 |
| 376 | 39-5094 | Skincare Specialists | -0.037 |
| 377 | 29-2011 | Medical and Clinical Laboratory Technologists | -0.042 |
| 378 | 19-4011 | Agricultural and Food Science Technicians | -0.049 |
| 379 | 29-1123 | Physical Therapists | -0.059 |
| 380 | 43-9031 | Desktop Publishers | -0.065 |
| 381 | 31-2011 | Occupational Therapy Assistants | -0.075 |
| 382 | 39-5092 | Manicurists and Pedicurists | -0.079 |
| 383 | 17-3029 | Engineering Technicians, Except Drafters, All Other | -0.079 |
| 384 | 29-1131 | Veterinarians | -0.084 |
| 385 | 29-2054 | Respiratory Therapy Technicians | -0.088 |
| 386 | 19-4051 | Nuclear Technicians | -0.091 |
| 387 | 19-4031 | Chemical Technicians | -0.094 |
| 388 | 29-9091 | Athletic Trainers | -0.095 |
| 389 | 41-2012 | Gaming Change Persons and Booth Cashiers | -0.104 |
| 390 | 41-2011 | Cashiers | -0.107 |
| 391 | 27-4012 | Broadcast Technicians | -0.111 |
| 392 | 25-4013 | Museum Technicians and Conservators | -0.111 |
| 393 | 39-3031 | Ushers, Lobby Attendants, and Ticket Takers | -0.111 |
| 394 | 49-1011 | First-Line Supervisors of Mechanics, Installers, and Repairers | -0.114 |
| 395 | 39-4031 | Morticians, Undertakers, and Funeral Directors | -0.114 |
| 396 | 53-6061 | Transportation Attendants, Except Flight Attendants | -0.118 |
| 397 | 27-4011 | Audio and Video Equipment Technicians | -0.120 |
| 398 | 17-1022 | Surveyors | -0.121 |
| 399 | 17-3025 | Environmental Engineering Technicians | -0.124 |
| 400 | 29-2051 | Dietetic Technicians | -0.128 |
| 401 | 29-2031 | Cardiovascular Technologists and Technicians | -0.131 |
| 402 | 29-2035 | Magnetic Resonance Imaging Technologists | -0.137 |
| 403 | 19-4091 | Environmental Science and Protection Technicians, Including Health | -0.143 |
| 404 | 29-1067 | Surgeons | -0.145 |
| 405 | 27-2023 | Umpires, Referees, and Other Sports Officials | -0.145 |
| 406 | 35-1012 | First-Line Supervisors of Food Preparation and Serving Workers | -0.150 |
| 407 | 33-2021 | Fire Inspectors and Investigators | -0.152 |
| 408 | 51-9061 | Inspectors, Testers, Sorters, Samplers, and Weighers | -0.156 |
| 409 | 39-4021 | Funeral Attendants | -0.158 |
| 410 | 11-9013 | Farmers, Ranchers, and Other Agricultural Managers | -0.158 |
| 411 | 27-1027 | Set and Exhibit Designers | -0.160 |
| 412 | 39-9011 | Childcare Workers | -0.167 |
| 413 | 33-3011 | Bailiffs | -0.171 |
| 414 | 35-1011 | Chefs and Head Cooks | -0.172 |
| 415 | 19-1032 | Foresters | -0.196 |
| 416 | 33-1012 | First-Line Supervisors of Police and Detectives | -0.196 |
| 417 | 29-1126 | Respiratory Therapists | -0.199 |
| 418 | 39-9021 | Personal Care Aides | -0.206 |
| 419 | 29-1024 | Prosthodontists | -0.214 |
| 420 | 27-1023 | Floral Designers | -0.219 |
| 421 | 39-3091 | Amusement and Recreation Attendants | -0.229 |
| 422 | 25-2059 | Special Education Teachers, All Other | -0.236 |
| 423 | 31-2021 | Physical Therapist Assistants | -0.247 |
| 424 | 31-9097 | Phlebotomists | -0.253 |
| 425 | 49-2091 | Avionics Technicians | -0.264 |
| 426 | 17-3027 | Mechanical Engineering Technicians | -0.274 |
| 427 | 29-2032 | Diagnostic Medical Sonographers | -0.274 |
| 428 | 31-1013 | Psychiatric Aides | -0.275 |
| 429 | 29-2099 | Health Technologists and Technicians, All Other | -0.281 |
| 430 | 29-2056 | Veterinary Technologists and Technicians | -0.282 |
| 431 | 29-2012 | Medical and Clinical Laboratory Technicians | -0.283 |
| 432 | 43-5051 | Postal Service Clerks | -0.284 |
| 433 | 17-3031 | Surveying and Mapping Technicians | -0.290 |
| 434 | 39-3092 | Costume Attendants | -0.296 |
| 435 | 29-2061 | Licensed Practical and Licensed Vocational Nurses | -0.298 |
| 436 | 35-3021 | Combined Food Preparation and Serving Workers, Including Fast Food | -0.308 |
| 437 | 33-9092 | Lifeguards, Ski Patrol, and Other Recreational Protective Service Workers | -0.316 |
| 438 | 29-1124 | Radiation Therapists | -0.316 |
| 439 | 53-6051 | Transportation Inspectors | -0.318 |
| 440 | 35-3022 | Counter Attendants, Cafeteria, Food Concession, and Coffee Shop | -0.321 |
| 441 | 29-1022 | Oral and Maxillofacial Surgeons | -0.330 |
| 442 | 51-6092 | Fabric and Apparel Patternmakers | -0.333 |
| 443 | 51-4012 | Computer Numerically Controlled Machine Tool Programmers, Metal and Plastic | -0.341 |
| 444 | 33-2022 | Forest Fire Inspectors and Prevention Specialists | -0.342 |
| 445 | 31-9091 | Dental Assistants | -0.346 |
| 446 | 29-1021 | Dentists, General | -0.348 |
| 447 | 29-2021 | Dental Hygienists | -0.352 |
| 448 | 43-5111 | Weighers, Measurers, Checkers, and Samplers, Recordkeeping | -0.353 |
| 449 | 31-2012 | Occupational Therapy Aides | -0.357 |
| 450 | 49-9061 | Camera and Photographic Equipment Repairers | -0.360 |
| 451 | 47-1011 | First-Line Supervisors of Construction Trades and Extraction Workers | -0.360 |
| 452 | 33-1011 | First-Line Supervisors of Correctional Officers | -0.361 |
| 453 | 53-2031 | Flight Attendants | -0.365 |
| 454 | 37-1011 | First-Line Supervisors of Housekeeping and Janitorial Workers | -0.367 |
| 455 | 27-4021 | Photographers | -0.382 |
| 456 | 33-9011 | Animal Control Workers | -0.390 |
| 457 | 35-2011 | Cooks, Fast Food | -0.391 |
| 458 | 33-9091 | Crossing Guards | -0.397 |
| 459 | 13-1074 | Farm Labor Contractors | -0.404 |
| 460 | 31-1011 | Home Health Aides | -0.405 |
| 461 | 29-2034 | Radiologic Technologists | -0.405 |
| 462 | 35-3011 | Bartenders | -0.411 |
| 463 | 33-9032 | Security Guards | -0.423 |
| 464 | 51-3092 | Food Batchmakers | -0.427 |
| 465 | 51-9081 | Dental Laboratory Technicians | -0.437 |
| 466 | 51-8013 | Power Plant Operators | -0.441 |
| 467 | 35-3041 | Food Servers, Nonrestaurant | -0.445 |
| 468 | 49-2011 | Computer, Automated Teller, and Office Machine Repairers | -0.454 |
| 469 | 35-3031 | Waiters and Waitresses | -0.469 |
| 470 | 51-3011 | Bakers | -0.474 |
| 471 | 37-1012 | First-Line Supervisors of Landscaping, Lawn Service, and Groundskeeping Workers | -0.486 |
| 472 | 39-5091 | Makeup Artists, Theatrical and Performance | -0.497 |
| 473 | 51-9151 | Photographic Process Workers and Processing Machine Operators | -0.498 |
| 474 | 39-5093 | Shampooers | -0.501 |
| 475 | 41-9012 | Models | -0.503 |
| 476 | 33-3031 | Fish and Game Wardens | -0.506 |
| 477 | 31-1014 | Nursing Assistants | -0.515 |
| 478 | 33-9093 | Transportation Security Screeners | -0.523 |
| 479 | 43-5081 | Stock Clerks and Order Fillers | -0.523 |
| 480 | 39-2021 | Nonfarm Animal Caretakers | -0.528 |
| 481 | 49-9062 | Medical Equipment Repairers | -0.537 |
| 482 | 29-2055 | Surgical Technologists | -0.542 |
| 483 | 39-5011 | Barbers | -0.543 |
| 484 | 51-9082 | Medical Appliance Technicians | -0.548 |
| 485 | 45-1011 | First-Line Supervisors of Farming, Fishing, and Forestry Workers | -0.558 |
| 486 | 35-2012 | Cooks, Institution and Cafeteria | -0.559 |
| 487 | 51-8031 | Water and Wastewater Treatment Plant and System Operators | -0.559 |
| 488 | 51-8091 | Chemical Plant and System Operators | -0.560 |
| 489 | 53-5031 | Ship Engineers | -0.567 |
| 490 | 35-2015 | Cooks, Short Order | -0.567 |
| 491 | 17-3024 | Electro-Mechanical Technicians | -0.568 |
| 492 | 51-8092 | Gas Plant Operators | -0.574 |
| 493 | 43-9071 | Office Machine Operators, Except Computer | -0.577 |
| 494 | 53-4031 | Railroad Conductors and Yardmasters | -0.579 |
| 495 | 51-8021 | Stationary Engineers and Boiler Operators | -0.580 |
| 496 | 49-9063 | Musical Instrument Repairers and Tuners | -0.584 |
| 497 | 31-9096 | Veterinary Assistants and Laboratory Animal Caretakers | -0.590 |
| 498 | 51-6052 | Tailors, Dressmakers, and Custom Sewers | -0.607 |
| 499 | 31-2022 | Physical Therapist Aides | -0.615 |
| 500 | 39-5012 | Hairdressers, Hairstylists, and Cosmetologists | -0.616 |
| 501 | 51-3093 | Food Cooking Machine Operators and Tenders | -0.619 |
| 502 | 49-9064 | Watch Repairers | -0.619 |
| 503 | 33-3041 | Parking Enforcement Workers | -0.627 |
| 504 | 27-4031 | Camera Operators, Television, Video, and Motion Picture | -0.629 |
| 505 | 51-2093 | Timing Device Assemblers and Adjusters | -0.635 |
| 506 | 33-1021 | First-Line Supervisors of Fire Fighting and Prevention Workers | -0.636 |
| 507 | 51-9071 | Jewelers and Precious Stone and Metal Workers | -0.637 |
| 508 | 35-2013 | Cooks, Private Household | -0.639 |
| 509 | 33-3051 | Police and Sheriff's Patrol Officers | -0.640 |
| 510 | 53-2012 | Commercial Pilots | -0.644 |
| 511 | 39-3021 | Motion Picture Projectionists | -0.649 |
| 512 | 51-8093 | Petroleum Pump System Operators, Refinery Operators, and Gaugers | -0.651 |
| 513 | 51-2023 | Electromechanical Equipment Assemblers | -0.652 |
| 514 | 39-4011 | Embalmers | -0.657 |
| 515 | 49-3052 | Motorcycle Mechanics | -0.666 |
| 516 | 33-3012 | Correctional Officers and Jailers | -0.669 |
| 517 | 49-2095 | Electrical and Electronics Repairers, Powerhouse, Substation, and Relay | -0.672 |
| 518 | 51-9011 | Chemical Equipment Operators and Tenders | -0.678 |
| 519 | 31-9011 | Massage Therapists | -0.682 |
| 520 | 49-2098 | Security and Fire Alarm Systems Installers | -0.685 |
| 521 | 51-4111 | Tool and Die Makers | -0.687 |
| 522 | 27-1013 | Fine Artists, Including Painters, Sculptors, and Illustrators | -0.698 |
| 523 | 29-2041 | Emergency Medical Technicians and Paramedics | -0.699 |
| 524 | 49-2094 | Electrical and Electronics Repairers, Commercial and Industrial Equipment | -0.703 |
| 525 | 45-4023 | Log Graders and Scalers | -0.703 |
| 526 | 43-9051 | Mail Clerks and Mail Machine Operators, Except Postal Service | -0.711 |
| 527 | 51-9141 | Semiconductor Processors | -0.713 |
| 528 | 53-3031 | Driver/Sales Workers | -0.722 |
| 529 | 37-2021 | Pest Control Workers | -0.724 |
| 530 | 53-3022 | Bus Drivers, School or Special Client | -0.729 |
| 531 | 39-2011 | Animal Trainers | -0.729 |
| 532 | 49-2097 | Electronic Home Entertainment Equipment Installers and Repairers | -0.730 |
| 533 | 27-1026 | Merchandise Displayers and Window Trimmers | -0.735 |
| 534 | 43-5071 | Shipping, Receiving, and Traffic Clerks | -0.738 |
| 535 | 49-2021 | Radio, Cellular, and Tower Equipment Installers and Repairers | -0.743 |
| 536 | 47-4041 | Hazardous Materials Removal Workers | -0.744 |
| 537 | 33-3052 | Transit and Railroad Police | -0.745 |
| 538 | 53-1011 | Aircraft Cargo Handling Supervisors | -0.751 |
| 539 | 51-2022 | Electrical and Electronic Equipment Assemblers | -0.752 |
| 540 | 53-2011 | Airline Pilots, Copilots, and Flight Engineers | -0.758 |
| 541 | 35-2014 | Cooks, Restaurant | -0.759 |
| 542 | 19-4093 | Forest and Conservation Technicians | -0.761 |
| 543 | 51-9194 | Etchers and Engravers | -0.775 |
| 544 | 31-9093 | Medical Equipment Preparers | -0.776 |
| 545 | 49-9012 | Control and Valve Installers and Repairers, Except Mechanical Door | -0.781 |
| 546 | 51-4062 | Patternmakers, Metal and Plastic | -0.788 |
| 547 | 51-3091 | Food and Tobacco Roasting, Baking, and Drying Machine Operators and Tenders | -0.790 |
| 548 | 53-5021 | Captains, Mates, and Pilots of Water Vessels | -0.793 |
| 549 | 49-9099 | Installation, Maintenance, and Repair Workers, All Other | -0.793 |
| 550 | 53-6011 | Bridge and Lock Tenders | -0.799 |
| 551 | 51-7031 | Model Makers, Wood | -0.802 |
| 552 | 51-8099 | Plant and System Operators, All Other | -0.811 |
| 553 | 49-3011 | Aircraft Mechanics and Service Technicians | -0.814 |
| 554 | 49-3091 | Bicycle Repairers | -0.823 |
| 555 | 51-4193 | Plating and Coating Machine Setters, Operators, and Tenders, Metal and Plastic | -0.828 |
| 556 | 51-2092 | Team Assemblers | -0.832 |
| 557 | 51-5112 | Printing Press Operators | -0.844 |
| 558 | 53-3011 | Ambulance Drivers and Attendants, Except Emergency Medical Technicians | -0.844 |
| 559 | 51-4081 | Multiple Machine Tool Setters, Operators, and Tenders, Metal and Plastic | -0.845 |
| 560 | 53-7011 | Conveyor Operators and Tenders | -0.846 |
| 561 | 51-9051 | Furnace, Kiln, Oven, Drier, and Kettle Operators and Tenders | -0.847 |
| 562 | 49-9091 | Coin, Vending, and Amusement Machine Servicers and Repairers | -0.852 |
| 563 | 51-4035 | Milling and Planing Machine Setters, Operators, and Tenders, Metal and Plastic | -0.867 |
| 564 | 51-4061 | Model Makers, Metal and Plastic | -0.868 |
| 565 | 51-2021 | Coil Winders, Tapers, and Finishers | -0.869 |
| 566 | 49-3092 | Recreational Vehicle Service Technicians | -0.869 |
| 567 | 49-2096 | Electronic Equipment Installers and Repairers, Motor Vehicles | -0.870 |
| 568 | 47-4099 | Construction and Related Workers, All Other | -0.871 |
| 569 | 53-7071 | Gas Compressor and Gas Pumping Station Operators | -0.873 |
| 570 | 47-5012 | Rotary Drill Operators, Oil and Gas | -0.880 |
| 571 | 51-4022 | Forging Machine Setters, Operators, and Tenders, Metal and Plastic | -0.881 |
| 572 | 27-2032 | Choreographers | -0.886 |
| 573 | 51-4032 | Drilling and Boring Machine Tool Setters, Operators, and Tenders, Metal and Plastic | -0.887 |
| 574 | 51-4011 | Computer-Controlled Machine Tool Operators, Metal and Plastic | -0.888 |
| 575 | 51-6011 | Laundry and Dry-Cleaning Workers | -0.889 |
| 576 | 49-9021 | Heating, Air Conditioning, and Refrigeration Mechanics and Installers | -0.896 |
| 577 | 49-2093 | Electrical and Electronics Installers and Repairers, Transportation Equipment | -0.910 |
| 578 | 53-4041 | Subway and Streetcar Operators | -0.913 |
| 579 | 51-5113 | Print Binding and Finishing Workers | -0.914 |
| 580 | 39-6011 | Baggage Porters and Bellhops | -0.914 |
| 581 | 53-4011 | Locomotive Engineers | -0.916 |
| 582 | 47-2111 | Electricians | -0.916 |
| 583 | 27-1012 | Craft Artists | -0.917 |
| 584 | 51-9196 | Paper Goods Machine Setters, Operators, and Tenders | -0.917 |
| 585 | 51-9083 | Ophthalmic Laboratory Technicians | -0.918 |
| 586 | 43-5021 | Couriers and Messengers | -0.922 |
| 587 | 51-6061 | Textile Bleaching and Dyeing Machine Operators and Tenders | -0.923 |
| 588 | 51-6091 | Extruding and Forming Machine Setters, Operators, and Tenders, Synthetic and Glass Fibers | -0.924 |
| 589 | 51-2011 | Aircraft Structure, Surfaces, Rigging, and Systems Assemblers | -0.927 |
| 590 | 47-4021 | Elevator Installers and Repairers | -0.928 |
| 591 | 49-9031 | Home Appliance Repairers | -0.928 |
| 592 | 37-3012 | Pesticide Handlers, Sprayers, and Applicators, Vegetation | -0.934 |
| 593 | 35-2021 | Food Preparation Workers | -0.934 |
| 594 | 51-4192 | Layout Workers, Metal and Plastic | -0.936 |
| 595 | 49-3051 | Motorboat Mechanics and Service Technicians | -0.938 |
| 596 | 53-3041 | Taxi Drivers and Chauffeurs | -0.938 |
| 597 | 49-2022 | Telecommunications Equipment Installers and Repairers, Except Line Installers | -0.939 |
| 598 | 51-9023 | Mixing and Blending Machine Setters, Operators, and Tenders | -0.941 |
| 599 | 45-2021 | Animal Breeders | -0.941 |
| 600 | 51-9031 | Cutters and Trimmers, Hand | -0.946 |
| 601 | 47-4091 | Segmental Pavers | -0.949 |
| 602 | 51-9123 | Painting, Coating, and Decorating Workers | -0.951 |
| 603 | 47-5031 | Explosives Workers, Ordnance Handling Experts, and Blasters | -0.955 |
| 604 | 49-3053 | Outdoor Power Equipment and Other Small Engine Mechanics | -0.961 |
| 605 | 43-5052 | Postal Service Mail Carriers | -0.967 |
| 606 | 53-7072 | Pump Operators, Except Wellhead Pumpers | -0.972 |
| 607 | 51-3021 | Butchers and Meat Cutters | -0.977 |
| 608 | 51-4041 | Machinists | -0.978 |
| 609 | 49-2092 | Electric Motor, Power Tool, and Related Repairers | -0.980 |
| 610 | 49-9081 | Wind Turbine Service Technicians | -0.984 |
| 611 | 49-9094 | Locksmiths and Safe Repairers | -0.991 |
| 612 | 47-5021 | Earth Drillers, Except Oil and Gas | -0.999 |
| 613 | 45-4011 | Forest and Conservation Workers | -1.000 |
| 614 | 51-6041 | Shoe and Leather Workers and Repairers | -1.007 |
| 615 | 43-5053 | Postal Service Mail Sorters, Processors, and Processing Machine Operators | -1.007 |
| 616 | 51-9111 | Packaging and Filling Machine Operators and Tenders | -1.011 |
| 617 | 53-7063 | Machine Feeders and Offbearers | -1.013 |
| 618 | 49-9043 | Maintenance Workers, Machinery | -1.015 |
| 619 | 51-6051 | Sewers, Hand | -1.019 |
| 620 | 53-4013 | Rail Yard Engineers, Dinkey Operators, and Hostlers | -1.021 |
| 621 | 51-3022 | Meat, Poultry, and Fish Cutters and Trimmers | -1.026 |
| 622 | 43-5041 | Meter Readers, Utilities | -1.027 |
| 623 | 37-2012 | Maids and Housekeeping Cleaners | -1.031 |
| 624 | 51-9193 | Cooling and Freezing Equipment Operators and Tenders | -1.031 |
| 625 | 49-3023 | Automotive Service Technicians and Mechanics | -1.031 |
| 626 | 45-2041 | Graders and Sorters, Agricultural Products | -1.032 |
| 627 | 47-2231 | Solar Photovoltaic Installers | -1.037 |
| 628 | 53-6021 | Parking Lot Attendants | -1.038 |
| 629 | 51-2031 | Engine and Other Machine Assemblers | -1.045 |
| 630 | 51-4122 | Welding, Soldering, and Brazing Machine Setters, Operators, and Tenders | -1.048 |
| 631 | 51-9012 | Separating, Filtering, Clarifying, Precipitating, and Still Machine Setters, Operators, and Tenders | -1.048 |
| 632 | 47-2042 | Floor Layers, Except Carpet, Wood, and Hard Tiles | -1.049 |
| 633 | 53-3021 | Bus Drivers, Transit and Intercity | -1.050 |
| 634 | 51-6042 | Shoe Machine Operators and Tenders | -1.052 |
| 635 | 49-9052 | Telecommunications Line Installers and Repairers | -1.054 |
| 636 | 49-3031 | Bus and Truck Mechanics and Diesel Engine Specialists | -1.058 |
| 637 | 47-5013 | Service Unit Operators, Oil, Gas, and Mining | -1.068 |
| 638 | 49-9071 | Maintenance and Repair Workers, General | -1.070 |
| 639 | 53-4012 | Locomotive Firers | -1.073 |
| 640 | 51-2041 | Structural Metal Fabricators and Fitters | -1.078 |
| 641 | 47-2132 | Insulation Workers, Mechanical | -1.080 |
| 642 | 49-9092 | Commercial Divers | -1.082 |
| 643 | 31-1015 | Orderlies | -1.088 |
| 644 | 45-2093 | Farmworkers, Farm, Ranch, and Aquacultural Animals | -1.089 |
| 645 | 37-2011 | Janitors and Cleaners, Except Maids and Housekeeping Cleaners | -1.093 |
| 646 | 51-4194 | Tool Grinders, Filers, and Sharpeners | -1.097 |
| 647 | 53-7073 | Wellhead Pumpers | -1.099 |
| 648 | 51-4072 | Molding, Coremaking, and Casting Machine Setters, Operators, and Tenders, Metal and Plastic | -1.108 |
| 649 | 53-6031 | Automotive and Watercraft Service Attendants | -1.108 |
| 650 | 51-6062 | Textile Cutting Machine Setters, Operators, and Tenders | -1.116 |
| 651 | 49-9051 | Electrical Power-Line Installers and Repairers | -1.119 |
| 652 | 47-2152 | Plumbers, Pipefitters, and Steamfitters | -1.120 |
| 653 | 47-2011 | Boilermakers | -1.122 |
| 654 | 51-9192 | Cleaning, Washing, and Metal Pickling Equipment Operators and Tenders | -1.124 |
| 655 | 51-4034 | Lathe and Turning Machine Tool Setters, Operators, and Tenders, Metal and Plastic | -1.126 |
| 656 | 53-7041 | Hoist and Winch Operators | -1.127 |
| 657 | 51-9021 | Crushing, Grinding, and Polishing Machine Setters, Operators, and Tenders | -1.136 |
| 658 | 51-9041 | Extruding, Forming, Pressing, and Compacting Machine Setters, Operators, and Tenders | -1.138 |
| 659 | 51-4052 | Pourers and Casters, Metal | -1.143 |
| 660 | 49-9097 | Signal and Track Switch Repairers | -1.148 |
| 661 | 51-4031 | Cutting, Punching, and Press Machine Setters, Operators, and Tenders, Metal and Plastic | -1.150 |
| 662 | 51-6063 | Textile Knitting and Weaving Machine Setters, Operators, and Tenders | -1.156 |
| 663 | 51-4191 | Heat Treating Equipment Setters, Operators, and Tenders, Metal and Plastic | -1.171 |
| 664 | 47-2073 | Operating Engineers and Other Construction Equipment Operators | -1.174 |
| 665 | 53-5022 | Motorboat Operators | -1.177 |
| 666 | 47-2151 | Pipelayers | -1.181 |
| 667 | 51-4023 | Rolling Machine Setters, Operators, and Tenders, Metal and Plastic | -1.186 |
| 668 | 53-7064 | Packers and Packagers, Hand | -1.193 |
| 669 | 51-4033 | Grinding, Lapping, Polishing, and Buffing Machine Tool Setters, Operators, and Tenders, Metal and Plastic | -1.194 |
| 670 | 47-2044 | Tile and Marble Setters | -1.196 |
| 671 | 51-9032 | Cutting and Slicing Machine Setters, Operators, and Tenders | -1.199 |
| 672 | 49-3021 | Automotive Body and Related Repairers | -1.199 |
| 673 | 53-3033 | Light Truck or Delivery Services Drivers | -1.200 |
| 674 | 51-7042 | Woodworking Machine Setters, Operators, and Tenders, Except Sawing | -1.204 |
| 675 | 51-7032 | Patternmakers, Wood | -1.208 |
| 676 | 49-9045 | Refractory Materials Repairers, Except Brickmasons | -1.209 |
| 677 | 49-9041 | Industrial Machinery Mechanics | -1.217 |
| 678 | 51-9022 | Grinding and Polishing Workers, Hand | -1.220 |
| 679 | 49-3022 | Automotive Glass Installers and Repairers | -1.235 |
| 680 | 51-7021 | Furniture Finishers | -1.238 |
| 681 | 49-3041 | Farm Equipment Mechanics and Service Technicians | -1.239 |
| 682 | 45-2091 | Agricultural Equipment Operators | -1.240 |
| 683 | 51-9195 | Molders, Shapers, and Casters, Except Metal and Plastic | -1.241 |
| 684 | 47-2211 | Sheet Metal Workers | -1.245 |
| 685 | 47-2031 | Carpenters | -1.245 |
| 686 | 53-7031 | Dredge Operators | -1.256 |
| 687 | 39-9031 | Fitness Trainers and Aerobics Instructors | -1.257 |
| 688 | 49-3042 | Mobile Heavy Equipment Mechanics, Except Engines | -1.259 |
| 689 | 47-4071 | Septic Tank Servicers and Sewer Pipe Cleaners | -1.260 |
| 690 | 53-7081 | Refuse and Recyclable Material Collectors | -1.263 |
| 691 | 51-7041 | Sawing Machine Setters, Operators, and Tenders, Wood | -1.265 |
| 692 | 47-2121 | Glaziers | -1.267 |
| 693 | 53-5011 | Sailors and Marine Oilers | -1.269 |
| 694 | 51-6031 | Sewing Machine Operators | -1.273 |
| 695 | 51-2091 | Fiberglass Laminators and Fabricators | -1.277 |
| 696 | 49-9096 | Riggers | -1.278 |
| 697 | 51-9121 | Coating, Painting, and Spraying Machine Setters, Operators, and Tenders | -1.279 |
| 698 | 47-4051 | Highway Maintenance Workers | -1.282 |
| 699 | 49-9011 | Mechanical Door Repairers | -1.282 |
| 700 | 47-5042 | Mine Cutting and Channeling Machine Operators | -1.283 |
| 701 | 49-9044 | Millwrights | -1.284 |
| 702 | 51-9191 | Adhesive Bonding Machine Operators and Tenders | -1.285 |
| 703 | 33-2011 | Firefighters | -1.287 |
| 704 | 27-2021 | Athletes and Sports Competitors | -1.290 |
| 705 | 51-4121 | Welders, Cutters, Solderers, and Brazers | -1.298 |
| 706 | 47-2071 | Paving, Surfacing, and Tamping Equipment Operators | -1.303 |
| 707 | 51-4051 | Metal-Refining Furnace Operators and Tenders | -1.304 |
| 708 | 53-4021 | Railroad Brake, Signal, and Switch Operators | -1.306 |
| 709 | 49-9098 | Helpers--Installation, Maintenance, and Repair Workers | -1.321 |
| 710 | 51-4021 | Extruding and Drawing Machine Setters, Operators, and Tenders, Metal and Plastic | -1.325 |
| 711 | 51-7011 | Cabinetmakers and Bench Carpenters | -1.327 |
| 712 | 47-2082 | Tapers | -1.345 |
| 713 | 53-7021 | Crane and Tower Operators | -1.352 |
| 714 | 45-4022 | Logging Equipment Operators | -1.355 |
| 715 | 47-5041 | Continuous Mining Machine Operators | -1.355 |
| 716 | 51-9198 | Helpers--Production Workers | -1.355 |
| 717 | 51-9199 | Production Workers, All Other | -1.363 |
| 718 | 53-7032 | Excavating and Loading Machine and Dragline Operators | -1.364 |
| 719 | 47-3012 | Helpers--Carpenters | -1.370 |
| 720 | 45-3021 | Hunters and Trappers | -1.374 |
| 721 | 47-2141 | Painters, Construction and Maintenance | -1.381 |
| 722 | 53-7033 | Loading Machine Operators, Underground Mining | -1.390 |
| 723 | 47-5011 | Derrick Operators, Oil and Gas | -1.391 |
| 724 | 53-3032 | Heavy and Tractor-Trailer Truck Drivers | -1.400 |
| 725 | 53-7121 | Tank Car, Truck, and Ship Loaders | -1.403 |
| 726 | 47-2061 | Construction Laborers | -1.405 |
| 727 | 51-6093 | Upholsterers | -1.405 |
| 728 | 47-2041 | Carpet Installers | -1.414 |
| 729 | 51-6064 | Textile Winding, Twisting, and Drawing Out Machine Setters, Operators, and Tenders | -1.417 |
| 730 | 49-3093 | Tire Repairers and Changers | -1.423 |
| 731 | 45-2092 | Farmworkers and Laborers, Crop, Nursery, and Greenhouse | -1.432 |
| 732 | 35-9011 | Dining Room and Cafeteria Attendants and Bartender Helpers | -1.433 |
| 733 | 49-9095 | Manufactured Building and Mobile Home Installers | -1.440 |
| 734 | 47-3013 | Helpers--Electricians | -1.445 |
| 735 | 47-2072 | Pile-Driver Operators | -1.455 |
| 736 | 47-3015 | Helpers--Pipelayers, Plumbers, Pipefitters, and Steamfitters | -1.462 |
| 737 | 51-9197 | Tire Builders | -1.463 |
| 738 | 49-9093 | Fabric Menders, Except Garment | -1.464 |
| 739 | 47-2022 | Stonemasons | -1.483 |
| 740 | 49-3043 | Rail Car Repairers | -1.485 |
| 741 | 47-2081 | Drywall and Ceiling Tile Installers | -1.496 |
| 742 | 47-4061 | Rail-Track Laying and Maintenance Equipment Operators | -1.497 |
| 743 | 53-7111 | Mine Shuttle Car Operators | -1.497 |
| 744 | 53-7062 | Laborers and Freight, Stock, and Material Movers, Hand | -1.498 |
| 745 | 47-5071 | Roustabouts, Oil and Gas | -1.500 |
| 746 | 47-2131 | Insulation Workers, Floor, Ceiling, and Wall | -1.510 |
| 747 | 45-3011 | Fishers and Related Fishing Workers | -1.511 |
| 748 | 53-7061 | Cleaners of Vehicles and Equipment | -1.535 |
| 749 | 51-3023 | Slaughterers and Meat Packers | -1.543 |
| 750 | 47-2181 | Roofers | -1.553 |
| 751 | 35-9021 | Dishwashers | -1.561 |
| 752 | 47-2142 | Paperhangers | -1.566 |
| 753 | 51-9122 | Painters, Transportation Equipment | -1.568 |
| 754 | 47-5081 | Helpers--Extraction Workers | -1.572 |
| 755 | 47-2043 | Floor Sanders and Finishers | -1.574 |
| 756 | 53-7051 | Industrial Truck and Tractor Operators | -1.575 |
| 757 | 47-2161 | Plasterers and Stucco Masons | -1.578 |
| 758 | 37-3013 | Tree Trimmers and Pruners | -1.581 |
| 759 | 47-5061 | Roof Bolters, Mining | -1.593 |
| 760 | 47-2021 | Brickmasons and Blockmasons | -1.607 |
| 761 | 47-5051 | Rock Splitters, Quarry | -1.614 |
| 762 | 51-4071 | Foundry Mold and Coremakers | -1.642 |
| 763 | 37-3011 | Landscaping and Groundskeeping Workers | -1.656 |
| 764 | 47-3016 | Helpers--Roofers | -1.668 |
| 765 | 47-4031 | Fence Erectors | -1.675 |
| 766 | 47-2051 | Cement Masons and Concrete Finishers | -1.678 |
| 767 | 47-2053 | Terrazzo Workers and Finishers | -1.685 |
| 768 | 47-2221 | Structural Iron and Steel Workers | -1.701 |
| 769 | 47-3014 | Helpers--Painters, Paperhangers, Plasterers, and Stucco Masons | -1.711 |
| 770 | 47-2171 | Reinforcing Iron and Rebar Workers | -1.781 |
| 771 | 45-4021 | Fallers | -1.791 |
| 772 | 27-2031 | Dancers | -1.793 |
| 773 | 47-3011 | Helpers--Brickmasons, Blockmasons, Stonemasons, and Tile and Marble Setters | -1.822 |
| 774 | 51-6021 | Pressers, Textile, Garment, and Related Materials | -1.854 |